\documentclass[twocolumn]{aastex631}

\usepackage{natbib}
\usepackage{multirow}
\usepackage{color}
\usepackage{bm}
\usepackage[normalem]{ulem}

\newcommand{\eg}{e.g., }
\newcommand{\ie}{i.e., }

\newcommand{\kms}{km~s$^{-1}$}

\def\gsim{\mathrel{\rlap{\lower 4pt \hbox{\hskip 1pt $\sim$}}\raise 1pt \hbox {$>$}}}
\def\lsim{\mathrel{\rlap{\lower 4pt \hbox{\hskip 1pt $\sim$}}\raise 1pt \hbox {$<$}}}

\newcommand{\teff}{T_{\rm eff}}

\shorttitle{Transition probabilities of NIR Ce III lines}
\shortauthors{N. Domoto et al.}

\begin{document}

\title{Transition probabilities of near-infrared Ce III lines from stellar spectra: applications to kilonovae}

\correspondingauthor{Nanae Domoto}
\email{n.domoto@astr.tohoku.ac.jp}

\author[0000-0002-7415-7954]{Nanae Domoto}
\affiliation{Astronomical Institute, Tohoku University, Aoba, Sendai 980-8578, Japan}

\author[0000-0003-0894-7824]{Jae-Joon Lee}
\affiliation{Korea Astronomy and Space Science Institute, 776 Daedeok-daero, Yuseong-gu, Daejeon, 34055, Republic of Korea}

\author[0000-0001-8253-6850]{Masaomi Tanaka}
\affiliation{Astronomical Institute, Tohoku University, Aoba, Sendai 980-8578, Japan}
\affiliation{Division for the Establishment of Frontier Sciences, Organization for Advanced Studies, Tohoku University, Sendai 980-8577, Japan}

\author[0000-0002-3808-7143]{Ho-Gyu Lee}
\affiliation{Korea Astronomy and Space Science Institute, 776 Daedeok-daero, Yuseong-gu, Daejeon, 34055, Republic of Korea}
\affiliation{Space Policy Research Center, Science and Technology Policy Institute, 370 Sicheong-daero, Sejong, 30147, Republic of Korea}

\author[0000-0002-8975-6829]{Wako Aoki}
\affiliation{National Astronomical Observatory, 2-21-1 Osawa, Mitaka, Tokyo 181-8588, Japan}
\affiliation{Astronomical Science Program, The Graduate University for Advanced Studies, SOKENDAI, 2-21-1 Osawa, Mitaka, Tokyo 181-8588, Japan}

\author[0000-0003-4656-0241]{Miho N. Ishigaki}
\affiliation{National Astronomical Observatory, 2-21-1 Osawa, Mitaka, Tokyo 181-8588, Japan}

\author[0000-0002-4759-7794]{Shinya Wanajo}
\affiliation{Max-Planck-Institut f\"{u}r Gravitationsphysik (Albert-Einstein-Institut), Am M\"{u}hlenberg 1, D-14476 Potsdam-Golm, Germany}
\affiliation{Interdisciplinary Theoretical and Mathematical Sciences Program (iTHEMS), RIKEN, Wako, Saitama 351-0198, Japan}

\author[0000-0002-5302-073X]{Daiji Kato}
\affiliation{National Institute for Fusion Science, 322-6 Oroshi-cho, Toki 509-5292, Japan}
\affiliation{Interdisciplinary Graduate School of Engineering Sciences, Kyushu University, Kasuga, Fukuoka 816-8580, Japan}

\author[0000-0002-2502-3730]{Kenta Hotokezaka}
\affiliation{Research Center for the Early Universe, Graduate School of Science, University of Tokyo, Bunkyo, Tokyo 113-0033, Japan}
\affiliation{Kavli IPMU (WPI), UTIAS, The University of Tokyo, Kashiwa, Chiba 277-8583, Japan}



\begin{abstract}
Kilonova spectra provide us with information of $r$-process nucleosynthesis in neutron star mergers.
However, it is still challenging to identify individual elements in the spectra mainly due to lack of experimentally accurate atomic data for heavy elements in the near-infrared wavelengths.
Recently, \citet{Domoto2022} proposed the absorption features around 14500 {\AA} in the observed spectra of GW170817/AT2017gfo as Ce III lines.
But they used theoretical transition probabilities ($gf$-values) whose accuracy is uncertain.
In this paper, we derive the astrophysical $gf$-values of the three Ce III lines, aiming at verification of this identification.
We model high resolution $H$-band spectra of four F-type supergiants showing the clear Ce III absorption features by assuming stellar parameters derived from optical spectra in literatures.
We also test the validity of the derived astrophysical $gf$-values by estimating Ce III abundances in Ap stars.
We find that the derived astrophysical $gf$-values of the Ce III lines are systematically lower by about 0.25 dex than those used in previous work of kilonovae, while they are still compatible within the uncertainty range.
By performing radiative transfer simulations of kilonovae with the derived $gf$-values, 
we find that the identification of Ce III as a source of the absorption features in the observed kilonova spectra still stands,
 even considering the uncertainties in the astrophysical $gf$-values.
This supports identification of Ce in the spectra of GW170817/AT2017gfo.
\end{abstract}



\section{Introduction}
\label{sec:intro}
Binary neutron star (NS) mergers have been considered as promising sites of $r$-process nucleosynthesis \citep[e.g.,][]{Eichler1989, Freiburghaus1999, Goriely2011, Korobkin2012, Wanajo2014}.
In 2017, associated with the detection of gravitational waves (GWs) from a NS merger \citep[GW170817,][]{Abbott2017a}, the electromagnetic counterpart AT2017gfo was observed \citep{Abbott2017b}.
The observed properties of AT2017gfo in ultraviolet, optical, and near-infrared (NIR) wavelengths are consistent with the theoretical expectation of a kilonova \citep{LiPaczynski1998, Metzger2010}, thermal emission from NS merger ejecta powered by radioactive decays of $r$-process nuclei \citep[e.g.,][]{Arcavi2017, Evans2017, Pian2017, Smartt2017, Utsumi2017, Valenti2017}.
This electromagnetic counterpart has provided us with evidence that NS mergers are sites of $r$-process nucleosynthesis \citep[e.g.,][]{Kasen2017, Perego2017, Shibata2017, Tanaka2017, Kawaguchi2018, Rosswog2018}.

To extract the detailed information of elements synthesized in NS merger ejecta, it is important to identify the individual elements.
However, element identification in the kilonova spectra remains challenging.
This is mainly due to lack of atomic data for heavy elements.
While detailed spectral studies require spectroscopically accurate atomic data, such data for heavy elements have been highly incomplete, especially for the NIR wavelengths.
So far, the absorption features around 8000 {\AA} in the photospheric spectra of AT2017gfo have been identified as Sr II \citep{Watson2019, Domoto2021, Gillanders2022}, although the same features may be caused by He I \citep{Perego2017, Tarumi2023}.

Recently, \citet{Domoto2022} systematically studied kilonova spectra over the entire wavelength range.
They identified important elements that produce strong transitions in kilonvoae, such as Ca, Sr, Y, Zr, Ba, La, and Ce.
Then, they constructed a hybrid line list combined with experimentally accurate data for important elements and complete theoretical data for weak transitions of the other elements \citep{Tanaka2020}.
By performing radiative transfer simulations, they found that the broad line features at $\sim$13000 {\AA} and 14500 {\AA} in the spectra of AT2017gfo can be explained by La III and Ce III, respectively.
However, this identification can be still a subject of discussion as they adopted the transition probabilities (\ie $gf$-values) of the La III and Ce III lines from theoretical calculations \citep{Tanaka2020} due to lack of experimental measurements.

The lack of experimental atomic data is problematic in any spectra for astrophysical objects, such as stars.
Compared with the optical region, line identification in stellar spectra in the NIR region is incomplete and still in progress \citep[e.g.,][]{Matsunaga2020}.
Also, as few experiments have been performed on heavy elements such as lanthanides, $gf$-values from theoretical calculations are commonly used even for the optical wavelengths.
In many cases, no or few data are provided for the NIR wavelengths \citep[e.g.,][]{DREAM1, DREAM2}.

In general, the accuracy of atomic calculations is difficult to assess without experimental measurements. 
It has been shown that the theoretical $gf$-values for given ions differ from the experimental values several times for strong transitions, and even more for weak transitions \citep[see, e.g., Appendix of][]{Domoto2022}.
The $gf$-values directly affect the absorption depths in the spectra, which influences derived parameters, \eg abundances.
In the case of kilonova model of \citet{Domoto2022}, the absorption features at the NIR wavelengths can disappear if the $gf$-values of the NIR lines are $\sim10$ times smaller than the values they adopted.
Therefore, it is important to measure the $gf$-values to firmly identify elements in kilonova spectra.

For the use in astrophysics, one can estimate the $gf$-values of atomic lines by using stellar spectra.
This is often called as ``astrophysical $gf$-values".
Such work has been performed for many metal lines in the NIR wavelengths using the spectra of well-observed stars, \eg the Sun \citep[e.g.,][]{MB1999}.
In fact, it has been shown that the astrophysical $gf$-values can be useful to extract elemental information from stars without laboratory experiments \citep{Hasselquist2016, Cunha2017}.

In this paper, we derive the astrophysical $gf$-values of three Ce III lines at the $H$-band (around 1.6 {\textmu}m) using high resolution stellar spectra.
This aims to verify the identification of Ce in the spectra of AT2017gfo.
Note that we cannot test the absorption by La III in the same way, because the La III lines at rest lie in the wavelength region of strong atmospheric absorption.
In Section \ref{sec:ce}, we describe the targeted Ce III lines in more detail.
We present the stellar spectra showing the Ce III line features in Section \ref{sec:data}, and derive the $gf$-values of the Ce III lines by modeling these spectra in Section \ref{sec:gf}.
In Section \ref{sec:kilonova}, we apply the derived $gf$-values of the Ce III lines to kilonova spectra by performing radiative transfer simulations.
Finally, we give a summary in Section \ref{sec:summary}.
We also show the validity of our results by estimating the Ce abundances with Ce III lines in Ap stars in Appendix \ref{sec:app}.
Throughout of the paper, line wavelengths are written as those in vacuum unless otherwise mentioned.

\begin{deluxetable*}{ccrrcccccccc}[ht]
\tablewidth{0pt}
\tablecaption{The targeted Ce III lines at $H$-band.}
\label{tab:CeIII}
\tablehead{
$\lambda_{\rm vac}$$^a$  & $\lambda_{\rm air}$$^b$ & $E_{\rm lower}$$^c$ & $E_{\rm upper}$$^d$ &  \multicolumn{4}{c}{Theoretical log $gf^e$} & \multicolumn{4}{c}{Astrophysical log $gf^f$}   \\
  ({\AA})       &    ({\AA})      &  (${\rm cm^{-1}}$)      & (${\rm cm^{-1}}$)  &  (1) & (2) & (3) & (4) & (5) & (6) & (7) & This work
}
\startdata
15851.880  &  15847.550  &    1528.32  &   7836.72  & $-$0.613 & $-$0.838 & $-$1.030 & $-$1.190 & $-$0.439 & $-$0.673 & $-$2.105   & $-$0.861 \\
15961.157  &  15956.797  &          0.00  &   6265.21  & $-$0.721 & $-$0.926 & $-$1.120 & $-$1.290 & $-$2.926 & $-$1.098 & $-$0.345   & $-$0.966 \\ 
16133.170  &  16128.763  &    3127.10  &   9325.51  & $-$0.509 & $-$0.722 & $-$0.920 & $-$1.070 & $-$0.429 & $-$1.950 & $-$0.920   & $-$0.702 \\
\enddata
\tablecomments{
$^a$ Vacuum transition wavelength. \\
$^b$ Air transition wavelength. \\
$^c$ Lower energy level. \\
$^d$ Upper energy level. \\
$^e$ $gf$-values from theoretical calculations (1: \citealp{Tanaka2020}, see also \citealp{Domoto2022}; 
	2: \citealp{WR1998}; 
	3: taken from \citealp{Hubrig2012} and \citealp{Chojnowski2019}. These values were computed in the same way as in \citealp{Biemont2002} and provided to \citealp{Hubrig2012} and \citealp{Chojnowski2019} as private communication; 
	4: \citealp{newgf}).
	For (1) and (4), theoretical $gf$-values are shown as they are, without calibration by considering the differences in the theoretical energy levels relative to the experimental energy levels. \\
$^f$ Astrophysical $gf$-values in the APOGEE line list (5: \citealp[Table 7 of][]{Shetrone2015}; 
	6: taken from \href{https://data.sdss.org/sas/dr13/apogee/spectro/redux/speclib/linelists/turbospec.20150714.atoms}{here} through the SDSS Science Archival Server (SAS), see \citealp{Holtzman2018}; 
	7: taken from \href{https://data.sdss.org/sas/dr17/apogee/spectro/speclib/linelists/turbospec/turbospec.20180901t20.atoms}{here} through the SDSS SAS, see \citealp{Smith2021}),
	and those derived in this work (Section \ref{sec:gf}).
}
\end{deluxetable*}

\section{Ce III lines at $H$-band}
\label{sec:ce}
We focus on the three Ce III lines at the $H$-band (Table \ref{tab:CeIII}).
It has been shown that these are the strongest lines of Ce III in the NIR region, which can produce broad absorption features in kilonova spectra \citep{Domoto2022}.
In fact, these lines have been detected in the spectra of Ap/Bp stars that exhibit enhanced Ce abundances \citep{Hubrig2012, Chojnowski2019, Tanaka2023}.

Although the transition wavelengths of these lines have been measured by experiments \citep{Johansson1972}, their transition probabilities are experimentally unknown.
Thus, \citet{Domoto2022} adopted the theoretical results of \citet{Tanaka2020}.
Other theoretical calculations are also available, \ie used for studies of optical stellar spectra \citep{WR1998, Biemont2002} and for application to kilonovae \citep{newgf}.
However, it is usually difficult to assess the accuracy of these calculations, and the $gf$-values vary over the range of $\sim0.5$ dex as shown in Table \ref{tab:CeIII}.

Recently, the APOGEE survey has observed many Galactic stars at the $H$-band with the spectral resolution of $R\sim22500$ \citep{apogee}.
To analyze the APOGEE spectra, the line lists for the $H$-band have been provided \citep{Shetrone2015, Holtzman2018, Smith2021}.
They collected the atomic data from theoretical and experimental studies, and adjusted the $gf$-values by fitting those to the spectra of the Sun and Arcturus.
It is shown that such astrophysical $gf$-values can improve the agreement between the observed spectra and the model spectra.
In fact, the APOGEE line lists include the three Ce III lines that we focus on (Table \ref{tab:CeIII}).
However, their astrophysical $gf$-values are quite uncertain, because there is no clear Ce III absorption in the spectra of the Sun and Arcturus.
As a result, the $gf$-values vary by an order of magnitude across the different versions of the APOGEE line lists.
This fact demonstrates that, if lines are not clearly detected in the spectra which is the case of Ce III lines for the Sun and Arcturus, it is difficult to derive the astrophysical $gf$-values even though the spectra are well characterized.
To derive $gf$-values of the Ce III lines, we select stars clearly showing these Ce III line features (Section \ref{sec:data}), although the uncertainties in stellar parameters can be larger than those of well-known stars such as the Sun and Arcturus (Section \ref{sec:unc}).

\begin{figure*}[th]
  \begin{center}
    \includegraphics[width=\linewidth]{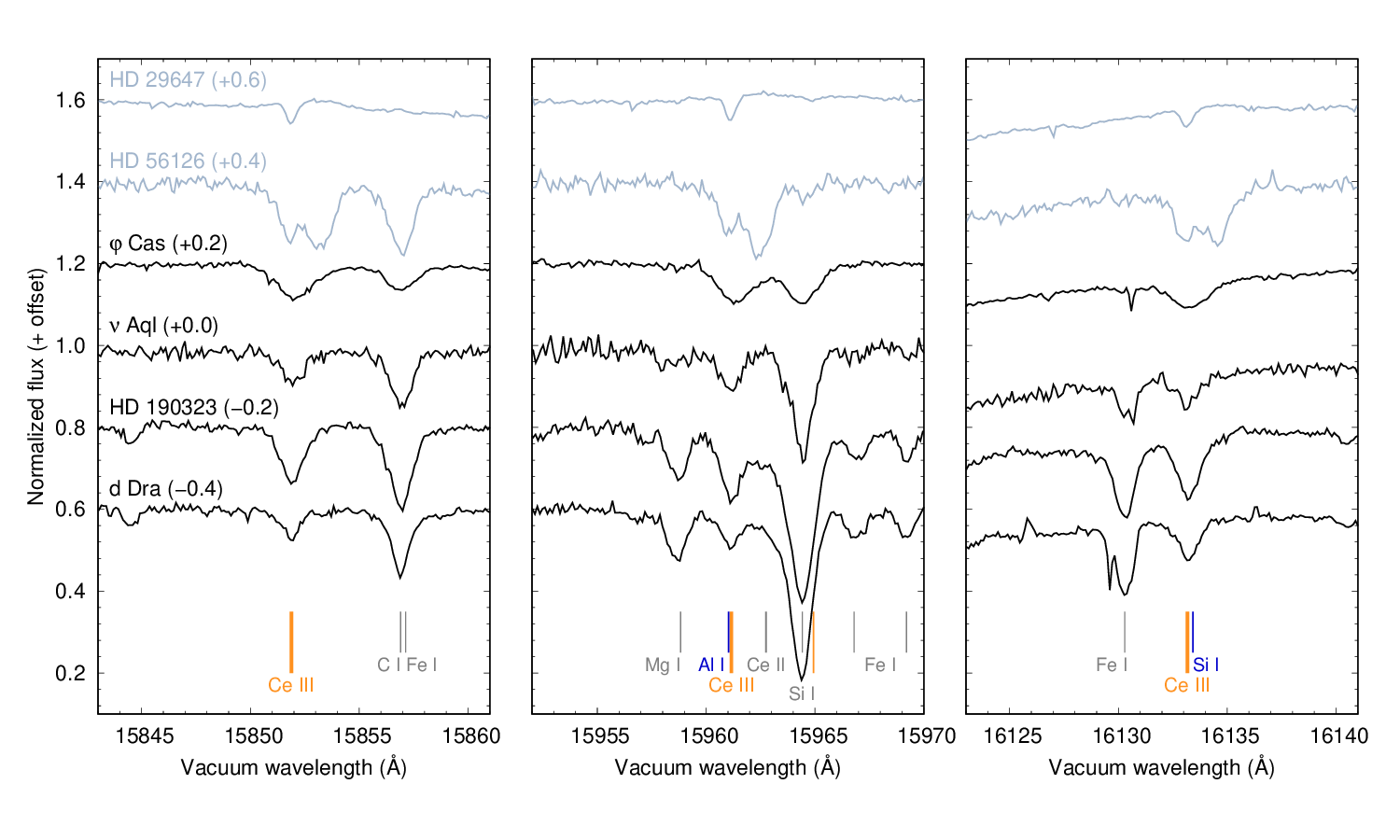}
\caption{
  \label{fig:obs_spec}
  The IGRINS spectra around the Ce III lines of the selected stars.
  Black and gray lines show the spectra for the objects used as the final sample and for the excluded objects, respectively (see the text).
  The positions of the three Ce III lines that we focus on are indicated by thick orange lines (Table \ref{tab:CeIII}), 
  while that of another Ce III line is also indicated by thin orange line (see Section \ref{sec:ce}).
  Al I and Si I lines that may contaminate the Ce III lines are labeled in blue.
  Other lines are also shown in gray.
}
\end{center}
\end{figure*}

We mention that near the three lines, there is another Ce III line at 15964.93 {\AA}.
According to the theoretical $gf$-values of this line ($-$1.272 by \citealp{Tanaka2020}, $-$1.660 by Biemont et al., see \citealp{Chojnowski2019}, or $-$1.690 by \citealp{newgf}), this line is weaker than the three Ce III lines.
In fact, it is found that the absorption at this wavelength is usually dominated by the strong Si I $\lambda$15964.41 {\AA} line (see Figure \ref{fig:obs_spec}).
Thus, we will not focus on this line in this work.

\section{Stellar spectra with Ce III lines}
\label{sec:data}
\subsection{Sample selection}
\label{sec:igrins}
Compared with the NIR wavelengths, the line list in the optical region is more comprehensive and accurate.
Ce abundances in many stars are indeed reported from optical observations, \eg in Hypatia Catalog for Solar neighborhoods \citep{Hinkel2014}, the catalogue of luminous stars \citep{Luck2014}, and the catalogue of chemically peculiar (CP) stars \citep{Ghazaryan2018}.
While not all the stars in these catalogues have Ce abundances, measurements are available for a good fraction of stars.
Thus, we will derive the astrophysical $gf$-values of the Ce III lines using the stars that (1) show the Ce III line features in the NIR $H$-band and (2) have known Ce abundances from optical spectra.

We obtain NIR $H$-band high resolution spectra from the spectral archive of the Immersion GRating INfrared Spectrometer \citep[IGRINS,][]{Park2014, Mace2016, Mace2018}.
IGRINS provides spectra spanning the full $H$ and $K$ bands with the spectral resolution of $R\sim45000$.
The Raw \& Reduced IGRINS Spectral Archive (RRISA)\footnote{https://igrinscontact.github.io/} contains the IGRINS archive for the period of July 2014--December 2021 on the 2.7 m Harlan J. Smith Telescope at McDonald Observatory, the Discovery Channel Telescope at Lowell Observatory, and the Gemini south telescope.
RRISA includes IGRINS data from more than 850 nights on-sky, resulting in 2800 unique target observations from over 3000 hours of the total science time.
RRISA also includes the IGRINS observations of $>1200$ distinct A0V standard stars.

RRISA provides reduced one-dimensional (1D) spectra processed using the IGRINS pipeline \citep{igrinspip}. 
The pipeline applies flat-fielding and flagging of bad pixels, and the 1D spectra are extracted using the optimal extraction of \citet{Horne1986}.
The wavelength solutions are derived using OH lines from sky spectra. 
In the $H$-band, the accuracy of wavelength solution is better than 0.01 {\AA} ($\sim$ 0.2 \kms).
RRISA further refines the wavelength of the pipeline product by improving the pixel alignment between the A0V star spectrum and the target spectrum using strong telluric sky lines between 16452--16458 {\AA} and 21740--21752 {\AA} in the $H$- and $K$-band, respectively.
Telluric absorption is corrected using the spectra of A0V stars observed closely in time and airmass with intrinsic features of A0V stars removed using a model spectrum of the Vega \citep{vega}.

We crossmatch the RRISA archive with the two stellar catalogues of measured abundances: \citet{Luck2014} and \citet{Ghazaryan2018}.
Not all entries in these catalogues have Ce abundances reported, and they are not used as we need abundance information to derive the $gf$-values (see below).
\citet{Luck2014} reports abundances for 451 luminous stars including Ce measurement for 448 stars.
RRISA contains spectra for 27 of them.
\citet{Ghazaryan2018} reports abundances of 429 CP stars with 121 stars of Ce measurement.
RRISA contains spectra of 8 stars.

\begin{figure}[th]
  \begin{center}
     \includegraphics[width=\linewidth]{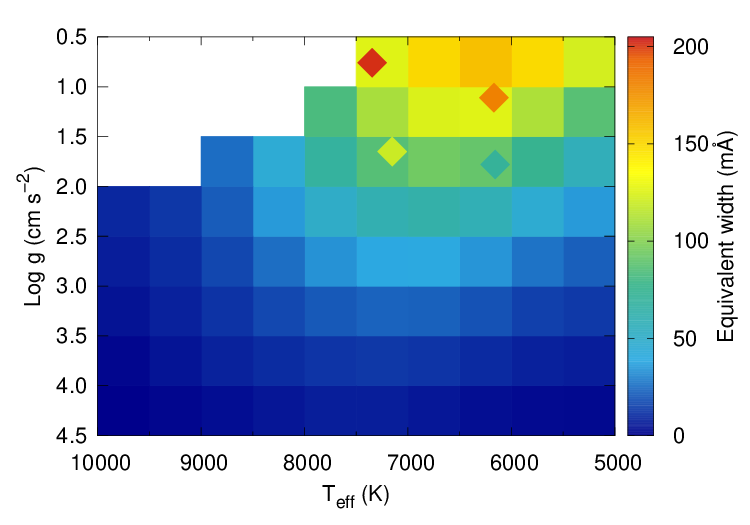}
\caption{
  \label{fig:ew}
  EW of the Ce III $\lambda$15961.157 {\AA} line in the model spectra (colormap) and in the observed spectra (diamonds).
  Note that the region without color (upper left) is the parameter space where no ATLAS model exists.
}
\end{center}
\end{figure}

For the matched data, we apply Doppler correction using strong C I, Si I, and Fe I lines.
Then, we visually check the absorption lines at the wavelengths of the Ce III lines.
We impose the detection of all three lines to ensure the origin of these lines.
As a result, six stars showing clear Ce III line features are selected: HD~29647, HD~56126, $\varphi$~Cas, $\nu$~Aql, HD~190323, and d~Dra.
The spectra of six stars around the regions of the Ce III lines are shown in Figure \ref{fig:obs_spec}.

While the six stars are selected above, we decide not to use two out of them: HD~29647 and HD~56126 (gray lines in Figure \ref{fig:obs_spec}).
This is because their atmospheric parameters are found to be rather uncertain, although they are listed in the catalogues.
For HD~29647, a HgMn star \citep{hd29647}, \citet{Adelman2001} have reported the atmospheric parameters, but they mentioned that the parameters are not very accurate.
For HD~56126, a post-AGB star \citep[e.g.,][]{hd56126dust}, atmospheric parameters seem to change depending on the epoch when spectra are taken \citep{DeSmedt2016, hd161796}.
\citet{hd161796} also pointed out the chemical depletion of neutron-capture elements in other post-AGB star, likely due to dust condensation.
Also, the IGRINS spectrum of HD~56126 shows double-peak like features, whose origins are unknown.

\begin{figure}[th]
  \begin{center}
    \includegraphics[width=\linewidth]{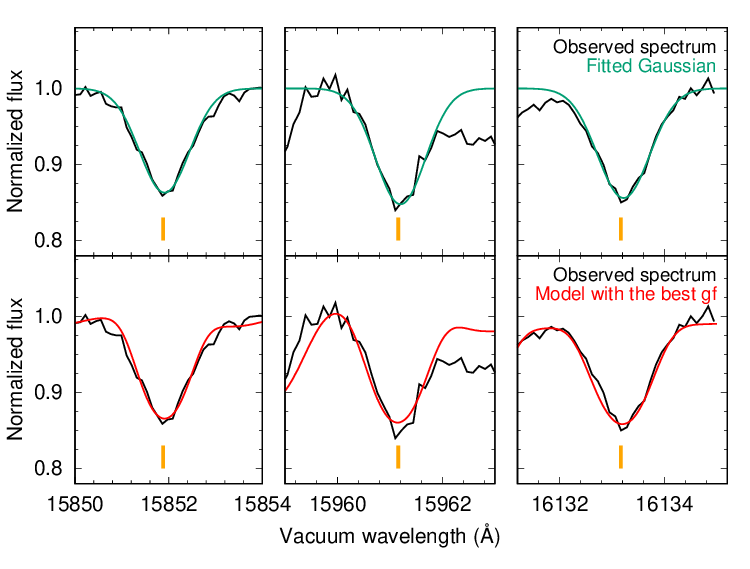}
\caption{
  \label{fig:ex}
  The Ce III lines of HD~190323 (black).
  In the top panels, the observed spectra are compared with the fitted Gaussian profile (green).
  In the bottom panels, the observed spectra are compared with the model spectra using the derived $gf$-values (red).
  Note that the observed spectrum at $\sim15963$ {\AA} of HD~190323 (the middle panels) may be affected by unknown lines.
}
\end{center}
\end{figure}

Finally, we use four stars as our samples (black lines in Figure \ref{fig:obs_spec}).
The S/N ratios of the spectra around the targeted Ce III lines are $>100$ for $\nu$~Aql and $>200$ for the others.
Since the Ce III $\lambda$15851.88 {\AA} and $\lambda$16133.170 {\AA} lines are located around the broad hydrogen features, the spectra are normalized by fitting the broad features as a pseudo-continuum.
After the normalization, we measure the equivalent widths (EWs) of the Ce III lines by Gaussian fitting (see the top panels of Figure \ref{fig:ex} as examples).
The measured EWs of the Ce III lines in each star are shown in Table \ref{tab:ew-gf}, which are used to derive those astrophysical $gf$-values in Section \ref{sec:gf}.

\begin{deluxetable*}{lccccccccccc}[ht]
\tablewidth{0pt}
\tablecaption{Adopted stellar parameters for the sample stars.}
\label{tab:param}
\tablehead{
Star   &   Spectral type &  Model &  $\teff$  &      log $g$       &  [Fe/H]  &  $\xi$   &  $v\sin i$ &  [Al/H]  &  [Si/H]  &  [Ce/H]  & $\sigma_{\rm[Ce/H]}$$^a$ \\
          &                         &            &   (K)      & (cm~s$^{-2}$)  &             &  (\kms) &  (\kms)     &             &             &             &
}
\startdata
$\varphi$~Cas &  F0Ia    & ATLAS   &  7347  &  0.76  &  $-$0.07  &  5.21  &  23.0$^*$           &  0.00$^b$  &  0.49  &  0.16  &  0.30\\
$\nu$~Aql        &  F2Ib    & ATLAS   &  7152  &  1.65  &       0.10  &  3.74  &  12.0$^{**}$        &  0.22          &  0.32  &  0.27  &  0.22 \\
HD~190323     &  F8Ia    & MARCS &  6169  &  1.11  &       0.12  &  4.31  &  15.0$^{\ddag}$  &  0.42          &  0.27  &  0.25  &  0.15\\
d~Dra              &  F8Ib-2 & MARCS &  6157  &  1.78  &  $-$0.10  &  3.93  &  10.0$^{\dag}$    &  0.16          &  0.06  &  0.27  &  0.24\\
\enddata
\tablecomments{
All the parameters are adopted from \citet{Luck2014}, but $v\sin i$ from \citet{phiCas} ($^*$), \citet{nuAql} ($^{**}$), \citet{dDra} ($^{\dag}$), and estimated in this work using the Ce III line because of no available measurement ($^{\ddag}$). \\
$^a$ Standard deviations of the Ce abundances measured in \citet{Luck2014} (see Section \ref{sec:unc}). \\
$^b$ The solar abundance \citep{Asplund2009} is assumed because it is not measured in \citet{Luck2014}. \\
}
\end{deluxetable*}

\subsection{Properties of sample stars}
\label{sec:prop}
The four stars in our final sample are all F-type supergiants (see Table \ref{tab:param}).
Our primary catalogue of \citet{Luck2014} consists primarily of (super)giants with $\teff\sim$ 4000--8000 K, and the Ce III lines are only seen in warmer supergiants of $\teff > 6000$ K.
To check the presence of the Ce III lines in wider parameter range of $\teff$ and surface gravity log $g$, we also crossmatch the RRISA with \citet{Hinkel2014}.
RRISA contains spectra of 132 sources out of 2685 stars from \citet{Hinkel2014}, primarily of nearby dwarf stars.
But none of the spectra show clear signature of the Ce III lines.
We further check RRSIA with $\teff$ and log $g$ from the PASTEL catalogue \citep{pastel}.
RRISA contains spectra of 168 stars with $\teff\sim5500$--8000 K, mostly dwarfs with some (super)giants.
Note that this catalogue does not provide abundance measurements and is not subject to our analysis of deriving the $gf$-values.
We find five stars showing the Ce III line features, which are the same four stars as our sample selected from \citet{Luck2014} and an additional F-type supergiant.

We briefly discuss the reason why the Ce III lines tend to appear in warm supergiants.
To see the behavior of the Ce III absorption in spectra, we generate model spectra for the wavelength regions around the Ce III lines, varying $\teff$ and log $g$.
We use Turbospectrum \citep{Plez2012} implemented through iSpec \citep{ispec1, ispec2} to perform spectral synthesis, assuming local thermodynamic equilibrium (LTE).
The parameter ranges are $\teff$ = 5000--10000 K and log $g$ = 0.5--4.5 dex with intervals of 500 K and 0.5 dex, respectively.
For simplicity, we fix the microturbulent velocity $\xi$ = 2 \kms, the solar abundance ratio \citep{Asplund2009}, and the projected rotational velocity $v\sin i$ = 10 \kms as typical values \citep[e.g.,][]{Matsunaga2020}, and assume the macroturbulent velocity $v_{\rm mac}=0$ \kms (see also Section \ref{sec:method}).
We use ATLAS, 1D plane-parallel atmosphere model \citep{atlas}.
For the atomic data, we use the APOGEE line list \citep[see Section \ref{sec:method}]{Holtzman2018, Smith2021} as a baseline, with the line list of the Ce III lines from \citet{Domoto2022}.
These choices do not affect the dependence on $\teff$ and log $g$ discussed here.

Since the dependence of the absorption on the atmospheric parameters is expected to be the same among the Ce III lines, we focus only on the Ce III $\lambda$15961.157 {\AA} line.
Note that this Ce III line may be contaminated by Al I $\lambda$15961.03 {\AA} line at lower $\teff$ and higher log $g$, although the Ce III line is dominant (Section \ref{sec:method}).
We measure the EWs of the Ce III line in the model spectra by Gaussian fitting as done in the observed spectra.
The measured EWs for different $\teff$ and log $g$ are shown in Figure \ref{fig:ew}.
It is seen that the EW of the Ce III line peaks at $\teff\sim$ 6000--7000 K and log $g <$ 2.
The peak $\teff$ can be understood as the region where the doubly ionized Ce is the most abundant under LTE.
Also, the line becomes strong toward low gravity, due to low-density and geometrically thick atmosphere \citep{Gray2005}.
A similar trend is observed for other lines of metal ions \citep{Matsunaga2020}.

We also show the observed EWs of the Ce III $\lambda$15961.157 {\AA} line for the four sample stars as shown with diamonds in Figure \ref{fig:ew}.
Although the Ce abundances are not exactly same as the solar abundances for these stars, it is seen that the observed trend is consistent with the trend in our model grid.
Also, models are consistent with the fact that only warm (F-type) supergiants show the Ce III lines but not normal dwarfs.

Of course, the behavior of the Ce III lines are also dependent on the Ce abundance.
If the Ce abundance is higher, the parameter range of $\teff$ and log $g$ showing the Ce III lines becomes wider.
For example, as shown in Figure \ref{fig:obs_spec}, HD~29647 shows Ce III absorption in the NIR wavelengths even under high temperature and strong surface gravity \citep[$\teff\sim12500$ K and log $g\sim4$,][]{Adelman2001}, because the Ce abundance is $\sim3$ dex higher than the solar value (see also Appendix \ref{sec:app}).
We note that CP stars are main sequences; supergiants used in our analysis do not show such anomaly in abundances of lanthanides.

\section{Astrophysical $gf$-values of the Ce III lines}
\label{sec:gf}
\subsection{Methods}
\label{sec:method}
In this section, we derive the astrophysical $gf$-values of the Ce III lines.
For each sample star, we perform spectral synthesis calculations with the fixed Ce abundances based on the measurements with the optical spectra, and then adjust $gf$-values.
We use Turbospectrum \citep{Plez2012} through iSpec \citep{ispec1, ispec2} to synthesize spectra as in Section \ref{sec:prop}.

We adopt atmospheric parameters ($\teff$, log $g$, metallicity, and $\xi$) derived from the analysis of optical spectra.
\citet{Luck2014} provided the stellar parameters of sample stars using MARCS models \citep{marcs}.
For the stars with $\teff$ higher than $\sim6300$ K, they also used ATLAS models \citep{atlas}.
In our sample, when a star has two parameter sets determined using both models, we use the atmospheric parameters that better reproduce the $H$-band IGRINS spectra.
Then, for spectral synthesis, we consistently use MARCS or ATLAS models used in the estimates of the adopted atmospheric parameters.
The adopted atmospheric parameters and models of sample stars are summarized in Table \ref{tab:param}.

We use the elemental abundances derived using the adopted atmospheric parameters in \citet{Luck2014}.
\citet{Luck2014} used observational data from multiple archives and provided abundances for each of the available data for a given star.
When multiple abundance sets are provided for our sample stars, we simply take the averages.
While all the elements are included in our calculations, only Ce, Al, and Si are important for our analysis:
Ce abundances directly affect the absorption depths of the Ce III lines, while Al and Si may also affect the Ce III absorption due to the contamination by Al I $\lambda$15961.03 {\AA} and Si I $\lambda$16133.42 {\AA} lines, respectively.
For the elements without measurements, we assume the solar abundances \citep{Asplund2009}.
For example, the Al abundance for $\varphi$~Cas is assumed as the solar abundance.
But this assumption does not affect the results because most Al is expected to be ionized in a high $\teff$ ($\sim7300$ K) and low log $g$ ($\sim0.8$ dex) atmosphere (see also Section \ref{sec:unc} and Table \ref{tab:unc}).
For all the samples, the contamination from the Al I and Si I lines for the Ce III absorption features are estimated to be $<10$\% and $<30$\%, respectively, by measuring the EWs of the lines in the model spectra without the Ce III lines.
The adopted abundances of important elements are also summarized in Table \ref{tab:param}.

It should be cautioned that, although the Ce abundances are fixed in our analysis, they are also subject to the uncertainties (see also Section \ref{sec:unc}).
The standard deviations of the Ce abundances measured in \citet{Luck2014} are also shown in Table \ref{tab:param}.
We confirmed that the absorption of Ce II lines at the $H$-band \citep{Cunha2017} is reasonably reproduced for HD~190323 and d~Dra, which confirms the validity of the abundances.
There is no or little Ce II absorption for $\varphi$~Cas and $\nu$~Aql with $\teff> 7000$ K.
Nevertheless, we emphasize that the Ce abundances in these two stars are rather robust, because they were derived using optical lines with experimental $gf$-values \citep{Luck2014}.

Spectral synthesis also needs to consider effects of line broadening.
It is usually difficult to separate the effects of $v\sin i$ and $v_{\rm mac}$ on the line profiles in F-type supergiants \citep[e.g.,][]{dDra}.
Here, we ignore $v_{\rm mac}$ and consider only $v\sin i$.
These parameters for $\varphi$~Cas, $\nu$~Aql, and d Dra are taken from literatures (see Table \ref{tab:param}).
We find that these choices reasonably reproduce the line profiles in the IGRINS spectra.
On the other hand, no measurement of $v\sin i$ for HD~190323 was found.
Thus, we estimate $v\sin i$ using the Ce III lines in the IGRINS spectra (see Table \ref{tab:param}).
Although this is not very accurate measurement, this assumption does not affect our results.

The line list for other elements is also an important factor in the estimate of $gf$-values.
This is because some Ce III lines are contaminated by Al I or Si I lines.
Thus, the $gf$-values of these lines may affect the estimates of the $gf$-values of the Ce III lines.
By comparing the $H$-band model spectra using the VALD \citep{Piskunov1995, Kupka1999, Ryabchikova2015} and the APOGEE line lists \citep{Holtzman2018, Smith2021}, we find that the APOGEE line list reproduces the observed features of the IGRINS spectra better.
Thus, we adopt the APOGEE line list for our analysis.
Note that, as iSpec implements the APOGEE line list of DR13/14 \citep{Holtzman2018}, we updated the line list using the newest DR16 line list \citep{Smith2021} except for hyperfine structure lines.

For the model spectra, we measure the EWs of the Ce III lines by Gaussian fitting as done in the observed spectra.
These EWs are compared with those of the observed spectra (Section \ref{sec:igrins}).
Comparison is repeated by changing the $gf$-values of the Ce III lines.
Finally, we adopt the $gf$-values that give the EWs in agreement with the EWs of the observed spectra.
Examples of the final model spectra are shown in the bottom panels of Figure \ref{fig:ex}.

\begin{figure*}[th]
  \begin{center}
    \includegraphics[width=\linewidth]{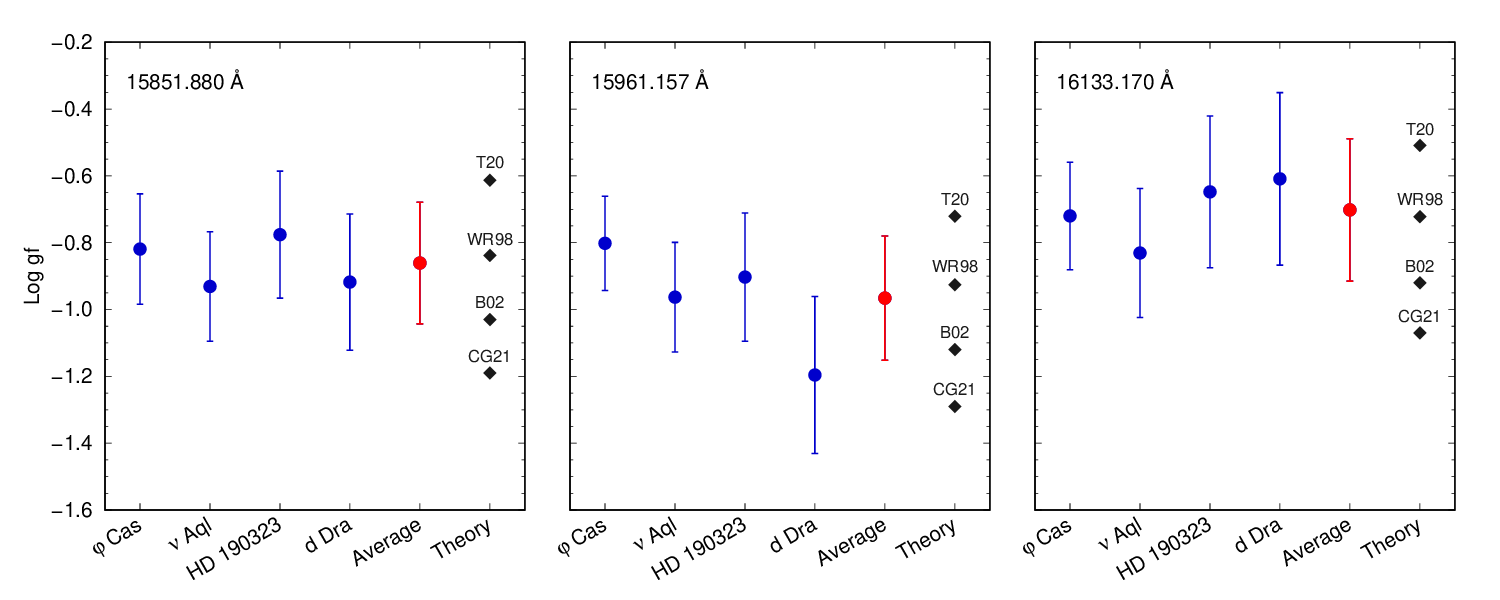}
\caption{
  \label{fig:loggf}
  Astrophysical $gf$-values of the Ce III lines derived in four stellar spectra (blue dots), and their average (red dots).
  Each panel shows the results of each transition as shown in the legend.
  Black diamonds show the theoretical $gf$-values 
  (T20: \citealp{Tanaka2020}; WR98: \citealp{WR1998}; B02: Biemont et al., see Table \ref{tab:CeIII}; CG21: \citealp{newgf}).
  Error bars indicate the estimated systematic uncertainties (see Section \ref{sec:unc} and Table \ref{tab:unc}).
  }
\end{center}
\end{figure*}
\begin{deluxetable*}{lcccccc}[ht]
\tablewidth{0pt}
\tablecaption{Observed EWs and derived $gf$-values for the three Ce III lines in four sample stars.}
\label{tab:ew-gf}
\tablehead{
Star   &     EW (m{\AA})   &      log $gf$            &     EW (m{\AA})   &      log $gf$            &     EW (m{\AA})   &      log $gf$  \\
          &    \multicolumn{2}{c}{15851.880 {\AA}} &    \multicolumn{2}{c}{15961.157 {\AA}} &    \multicolumn{2}{c}{16133.170 {\AA}}
}
\startdata
$\varphi$~Cas &  184  & $-$0.82  &  201  &  $-$0.80  &  166  &  $-$0.72  \\
$\nu$~Aql        &  102  & $-$0.93  &  120  &  $-$0.96  &    96  &  $-$0.83  \\
HD~190323     &  183  & $-$0.78  &  182  &  $-$0.90  &  180  &  $-$0.65  \\
d~Dra               &    81  & $-$0.92  &    71  &  $-$1.20  &  113  &  $-$0.61 \\
\enddata
\end{deluxetable*}
\begin{deluxetable*}{lcccccccc}[ht]
\tablewidth{0pt}
\tablecaption{Systematic uncertainties in the astrophysical $gf$-values of the Ce III lines for the fixed Ce abundances in each star.}
\label{tab:unc}
\tablehead{
Star   &  $\lambda_{\rm vac}$ & $\Delta \teff$ & $\Delta$log $g$ &  $\Delta$[Fe/H]  &  $\Delta\xi$  &  $\Delta$[Al/H]  &  $\Delta$[Si/H] & $\Delta_{\rm sys}$ \\
          &            ({\AA})             &   (+200 K)      &   (+0.3 dex)       &   (+0.3 dex)         &  (+0.5 \kms) & (+0.5 dex)   & (+0.3 dex) & 
}
\startdata
$\varphi$~Cas &  15851.880    &    0.146   &  0.067  &  0.035  &  $-$0.018  &        -        &         -        &  0.165 \\
                        &  15961.157    &    0.105   &  0.085  &  0.029  &  $-$0.030  &  $-$0.002 &         -        &  0.141  \\
                        &  16133.170    &    0.139   &  0.070  &  0.034  &  $-$0.010  &        -        &  $-$0.022  &  0.161  \\ \hline
$\nu$~Aql        &  15851.880    &    0.048   &  0.153  &  0.028  &  $-$0.017  &        -        &         -        &  0.164  \\
                        &  15961.157    &    0.054   &  0.151  &  0.025  &  $-$0.019  &  $-$0.015 &         -        &  0.164  \\
                        &  16133.170    &    0.067   &  0.151  &  0.046  &  $-$0.014  &        -        &  $-$0.088  &  0.193  \\ \hline
HD~190323     &  15851.880    &    0.000   &  0.176  &  0.066  &  $-$0.030  &        -        &         -        &  0.190   \\
                        &  15961.157    &    0.003   &  0.179  &  0.057  &  $-$0.021  &  $-$0.031  &         -       &  0.192  \\
                        &  16133.170    &    0.018   &  0.175  &  0.100  &  $-$0.026  &        -         &  $-$0.099 &  0.227  \\ \hline
d~Dra              &  15851.880    & $-$0.015 &  0.196  &  0.055  &  $-$0.011  &        -         &         -        &  0.204  \\
                        &  15961.157    & $-$0.002 &  0.194  &  0.056  &  $-$0.005  &  $-$0.121  &         -        &  0.235  \\
                        &  16133.170    &    0.014   &  0.194  &  0.081  &  $-$0.014  &        -         &  $-$0.149  &  0.258  \\
\enddata
\end{deluxetable*}

\subsection{Results}
\label{sec:result}
The estimated $gf$-values of the three Ce III lines using the spectra of four stars are shown by blue dots in Figure \ref{fig:loggf}, and listed in Table \ref{tab:ew-gf}.
As our final astrophysical $gf$-values of each Ce III line, we take the averages of those derived for each star.
They are plotted as red dots in each panel of Figure \ref{fig:loggf}, and listed in Table \ref{tab:CeIII}.
The validity of our final $gf$-values is also discussed in Appendix \ref{sec:app}.

The $gf$-values of the Ce III lines from available theoretical calculations are also shown by black diamonds in Figure \ref{fig:loggf} (see Table \ref{tab:CeIII}).
All of our $gf$-values are consistently smaller than those of \citet{Tanaka2020}, which were used for the kilonova model of \citet{Domoto2022}.
This suggests that the strength of absorption could have been overestimated in the kilonova model of \citet{Domoto2022} (see Section \ref{sec:kilonova}).
On the other hand, it is seen that the estimated final $gf$-values are the closest to those of \citet{WR1998}, while higher than those of \citet{Biemont2002} and \citet{newgf}.
Overall, our $gf$-values of the Ce III lines, which are independently estimated from theoretical calculations, broadly agree with the available theoretical calculations within the margin of the estimated uncertainties discussed in Section \ref{sec:unc}.

\subsection{Uncertainties in $gf$-values}
\label{sec:unc}
Uncertainties in the astrophysical $gf$-values can be evaluated by considering the uncertainties in the stellar parameters of each star.
The systematic uncertainty of stellar parameters can generally be caused by difference in the analysis methods.
Even with common atmospheric parameters and line lists, there remain variety in the abundances determined by different codes and methods \citep{Hinkel2016} as well as different spectral data used in the analysis \citep{Luck2014}.
All of these factors can affect the depth of absorption lines, which leads to uncertainties in the astrophysical $gf$-values derived for the fixed Ce abundances.

We estimate the systematic uncertainties in our astrophysical $gf$-values by varying stellar parameters of each star.
To take uncertainties of stellar parameters into account, we vary $\teff$ by $\pm$200 K, log $g$ by $\pm$0.3 dex, [Fe/H] by $\pm$0.3 dex, and $\xi$ by $\pm$0.5 \kms.
Also, we vary [Al/H] and [Si/H] by $\pm$0.5 and $\pm$0.3 dex, respectively, which affect the level of the contamination.
These ranges are typical systematic uncertainties caused by different codes and methods used to determine stellar parameters \citep{Hinkel2016}.
While \citet{Hinkel2016} used dwarfs to show these uncertainty ranges, the uncertainties for supergiants are similar to those in dwarfs \citep{Luck2014}.

Table \ref{tab:unc} summarizes the estimated systematic uncertainties in the astrophysical $gf$-values of the Ce III lines.
Only the results for the positive shifts are shown, but the results for the negative shifts are smaller than the uncertainties given in Table \ref{tab:unc} \citep[see also Table 3 of ][]{Cunha2017}.
We find that the largest uncertainties are caused by $\teff$ or log $g$.
The systematic uncertainties in total in each $gf$-value for each star ($\Delta_{\rm sys}$) are estimated by taking root sum squares \citep{Cunha2017}.
Note that the values of $\Delta_{\rm sys}$ estimated here are the upper limits of the uncertainties:
all the stellar parameters are not independent, and the uncertainties caused by each parameter may be cancelled among parameters \citep{Cunha2017}.
In most cases, the total systematic uncertainties are smaller than 0.2 dex, and up to $\sim0.26$ dex.
These systematic uncertainties are shown as the error bars for the blue dots in Figure \ref{fig:loggf}.

Finally, we assign the systematic uncertainties of the average astrophysical $gf$-values by taking root mean squares of the $\Delta_{\rm sys}$ in four stars.
The systematic uncertainties in the $gf$-values for $\lambda$15851.880 {\AA}, $\lambda$15961.157 {\AA}, and $\lambda$16133.170 {\AA} lines are 0.182, 0.186, and 0.213, respectively (shown as the error bars for the red dots in Figure \ref{fig:loggf}).
It is worth noting that our astrophysical $gf$-values are broadly agree with available theoretical calculations, considering the systematic uncertainties.

Here we did not show the uncertainties in continuum levels of the observed spectra in $\Delta_{\rm sys}$.
They may affect the observed EWs and in turn astrophysical $gf$-values.
However, the uncertainties in continuum leveles and their contributions to $\Delta_{\rm sys}$ are estimated to be $\lesssim$ 1\%, which are smaller than the uncertainties from most stellar parameters \citep[see also Table 3 of][]{Cunha2017}.

Note that, although we fix the Ce abundances of each star to derive astrophysical $gf$-values, the Ce abundances are still uncertain within the range of $\lesssim0.3$ dex (standard deviations, Section \ref{sec:method}).
The Ce abundances scale the resultant $gf$-values:
for example, if the Ce abundance is altered by the standard deviation (Table \ref{tab:param}) to the positive direction, the derived $gf$-values are also affected by the same degree to the negative direction.
Furthermore, it should be noted that the Ce abundances in \citet{Luck2014} as well as the astrophysical $gf$-values here were derived under the assumption of LTE.
Non-LTE treatment may affect the abundances of heavy elements \citep[e.g.,][]{Mashonkina2005}, especially in supergiants.
This may result in systematic changes in the derived $gf$-values.
Nevertheless, the fact that the derived $gf$-values are consistent among samples suggests that the Ce abundance of each star adopted in this work is reasonable.
Independent measurements of stellar parameters may be able to reduce the uncertainties in the astrophysical $gf$-values, but this is beyond the scope of this work.
The uncertainties present here are found not to significantly affect the final results on kilonova spectra, as discussed in Section \ref{sec:kilonova}.

\begin{figure}[th]
  \begin{center}
    \includegraphics[width=\linewidth]{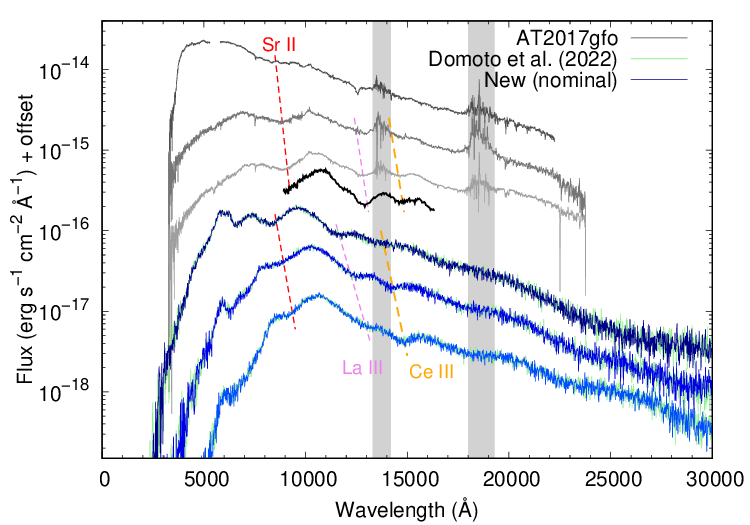}
\caption{
  \label{fig:spec}
  Comparison between the synthetic spectra (blue) and the observed spectra of AT2017gfo taken with the VLT \citep[gray,][]{Pian2017, Smartt2017} at $t=1.5$, 2.5, and 3.5 days after the merger (dark to light colors).
  The spectrum taken with the HST at $t=4.5$ days after the merger \citep{Tanvir2017} is also shown in black line.
  The green curves are the synthetic spectra of \citet[see their Figure 8]{Domoto2022}.
  Spectra are vertically shifted for visualization.
  Gray shade shows the regions of strong atmospheric absorption.
  }
\end{center}
\end{figure}
\begin{figure}[th]
  \begin{center}
    \includegraphics[width=\linewidth]{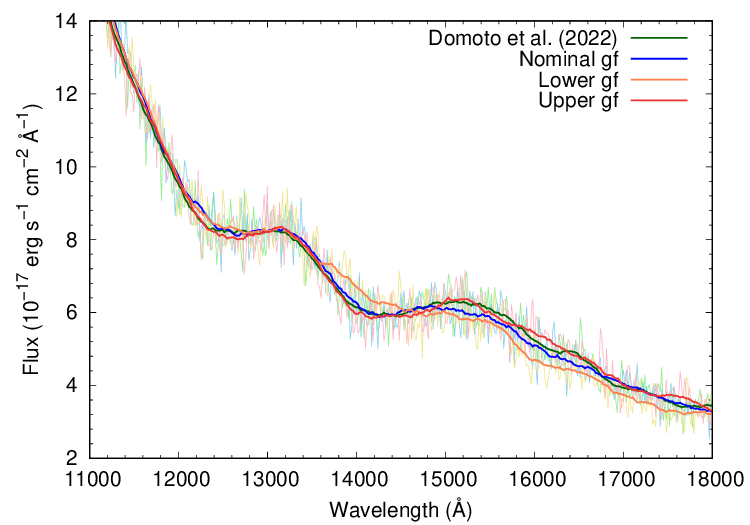}
\caption{
  \label{fig:comp}
  Enlarged view of the synthetic spectra for the NIR region at $t=2.5$ days.
  Blue curves show the synthetic spectra using the derived $gf$-values, and green curves show those of \citet{Domoto2022}.
  Orange and red curves show the synthetic spectra when using the lower and upper $gf$-values, which are considering the sum of the uncertainties negatively and positively, respectively (see the text).
  The curves in light colors show the original results, while those in dark colors show smoothed spectra for visualization.
  }
\end{center}
\end{figure}

\section{Applications to kilonova spectra}
\label{sec:kilonova}
We apply the derived $gf$-values of the Ce III lines to calculate kilonova spectra.
To calculate synthetic spectra of kilonovae, we use a wavelength-dependent radiative transfer simulation code \citep{TH2013, Tanaka2014, Tanaka2017, Tanaka2018, Kawaguchi2018, Kawaguchi2020}.
The photon transfer is calculated by the Monte Carlo method.
The setup of simulation is identical to that in \citet{Domoto2022}, but we adopt the final astrophysical $gf$-values of the three Ce III lines derived in Section \ref{sec:gf}.
For more details of the simulation, we refer the readers to \citet{Domoto2022}.

Figure \ref{fig:spec} shows the comparison between our new synthetic spectra (blue) and the observed spectra of AT2017gfo taken with the Very Large Telescope (VLT) (gray) at $t = 1.5$, 2.5, and 3.5 days after the merger \citep{Pian2017, Smartt2017}.
The observed spectra at $t=4.5$ days after the merger taken with the Hubble Space Telescope (HST), which are not affected by telluric absorption, is also shown by a black line \citep{Tanvir2017}.
We find that the absorption features around 14500 {\AA} are caused by the Ce III lines in the new spectra.
These features are blueshifted according to the velocity of the line-forming region at the NIR wavelengths, \eg $v\sim0.1\ c$ at $t=2.5$ days.
In fact, the features caused by the Ce III lines in the new synthetic spectra are almost the same as those in \citet[green]{Domoto2022}, which used the $gf$-values of \citet{Tanaka2020} (see Figure \ref{fig:comp} for the enlarged view at $t=2.5$ days).
This is consistent with the observed features in the spectra of AT2017gfo.

To see the effects of the uncertainties in the $gf$-values to the spectra, we also perform the same simulations by varying the $gf$-values of the Ce III lines.
Orange and red curves in Figure \ref{fig:comp} show the results when using the lower and upper $gf$-values, respectively.
Here, for the lower (upper) values, we take the averages of the estimated $gf$-values for each star negatively (positively) shifed by $\sigma_{\rm [Ce/H]}$, and then further shift the averages by $\Delta_{\rm sys}$ negatively (positively).
We find that the changes in fluxes around 14500 {\AA} are $\sim10$\%.
The absorption features by Ce III are still seen even using the lower $gf$-values (orange curves), although the depth of absorption is weakened.

When using the final astrophysical $gf$-values, the Sobolev optical depths of the Ce III lines for the line forming region ($\rho\sim10^{-14}$ g~cm$^{-3}$ and $T\sim5000$ K, see \citealp{Domoto2022}) are $\sim4$.
The lower $gf$-values of the Ce III lines are smaller than the nominal values by a factor of $\sim2.7$.
Thus, even by adopting the lower $gf$-values, the Sobolev optical depths of the Ce III lines are still larger 1.
This confirms the identification of Ce in the spectra of AT2017gfo.

The changes in $gf$-values of the strong lines affect not only absorption but also emission in the spectra.
A stronger absorption tends to cause a stronger emission in P-Cygni type line profile.
As shown in Figure \ref{fig:comp}, the increase of the $gf$-values results in the increase of the flux at emission component around the rest wavelengths, and vice versa.
Note that, for particularly large $gf$-values (red and green lines), the optical depths of the Ce III lines are large enough that the depth of the absorption is saturated, while the effect in the emission component is still apparent.

\section{Summary}
\label{sec:summary}
We have derived the astrophysical $gf$-values of the three Ce III lines at the $H$-band using the IGRINS spectra of four F-type supergiants.
The derived $gf$-values are systematically lower by about 0.25 dex than those used in \citet{Domoto2022}, but they are broadly agree with available theoretical calculations within the estimated uncertainties.
Using the derived astrophysical $gf$-values of the Ce III lines, we have performed radiative transfer simulations of kilonovae.
We have found that the Ce III lines with the new $gf$-values produce the absorption features around 14500 {\AA} in kilonova spectra, even considering the uncertainties.
This supports the identification of Ce in the observed spectra of AT2017gfo \citep{Domoto2022}.

We have also shown that the F-type supergiants with near solar metallicity show strong absorption of Ce III.
To our knowledge, this is the first report of the detection of the Ce III lines at the $H$-band in stars with near solar abundances.
As the line blending is not severe in the NIR spectra compared with the optical spectra, those lines may also be useful for stellar spectroscopic studies.

\begin{acknowledgments}
We thank R. E. Luck for providing his detailed abundance data, and K. Kawaguchi for valuable discussions.
This work used the Immersion Grating Infrared Spectrometer (IGRINS) that was developed under a collaboration between the University of Texas at Austin and the Korea Astronomy and Space Science Institute (KASI) with the financial support of the Mt. Cuba Astronomical Foundation, of the US National Science Foundation under grants AST-1229522 and AST-1702267, of the McDonald Observatory of the University of Texas at Austin, of the Korean GMT Project of KASI, and Gemini Observatory.
Part of the observations in this work were also carried out within the framework of Subaru-Gemini time exchange program which is operated by the National Astronomical Observatory in Japan.
We are honored and grateful for the opportunity of observing the Universe from Maunakea, which has the cultural, historical and natural significance in Hawaii.
Part of numerical simulations presented in this paper were carried out on Cray XC50 at Center for Computational Astrophysics, National Astronomical Observatory of Japan.
This work was supported by JST FOREST Program (Grant Number JPMJFR212Y, JPMJFR2136), NIFS Collaborative Research Program (NIFS22KIIF005), the Grant-in-Aid for JSPS Fellows (22KJ0317), and the Grant-in-Aid for Scientific Research from JSPS (19H00694, 20H00158, 21H04997, 23H00127, 23H04891, 23H04894, 23H05432).
N.D. acknowledges support from Graduate Program on Physics for the Universe (GP-PU) at Tohoku University.

\end{acknowledgments}

\appendix
\section{Abundance estimates in Ap stars as a test of astrophysical $gf$-values}
\label{sec:app}
We assess the validity of our astrophysical $gf$-values by adopting them to spectra of Ap stars.
Ap stars are a kind of CP stars that exhibit extremely high metal abundances.
Thanks to the high abundances, Ap stars show clear Ce III absorption lines that we focus on.
However, it is also known that they show inconsistent abundances between singly and doubly ionized rare-earth elements (REE), called REE anomaly \citep{Ryabchikova2004, Ryabchikova2017}.
Even if Ce abundances have been measured from Ce II lines, it does not necessarily represent the Ce abundances.
Therefore, these stars cannot be used to derive $gf$-values in Section \ref{sec:gf}.

Here, we estimate the Ce abundances from Ce III lines (hereafter Ce III abundances) in Ap stars using the derived astrophysical $gf$-values.
We use three Ap stars, HD~101065 (Przybylski's Star), HD~201601 ($\gamma$ Equ), and HD~24712.
These stars have quite different log $g$ from the supergiants used to derive the $gf$-values.
Thus, the abundance estimates serve as independent assessment of the derived $gf$-values.

The spectrum of HD~101065 was taken with Gemini south/IGRINS on UT 2022 December 18 in two ABBA sequences with each exposure time of 33 sec.
An A0V star (HIP~55019) was also observed for the telluric standard.
Those IGRINS spectra are reduced through the pipeline \citep{igrinspip}, as described in Section \ref{sec:igrins}.
The 1D spectra are normalized by continuum fitting.

The spectra of HD~201601 and HD~24712 were taken with the Gemini Near-Infrared Spectrograph \citep[GNIRS,][]{gnirs2, gnirs1} at Gemini north telescope on UT 2022 July 17 and August 18, respectively.
We use 110.5 l/mm grating and long camera with a 0.1'' slit, which gives the spectral resolution of $R\sim$17800.
Central wavelength is set to be 1.6 {\textmu}m to cover the three Ce III lines.
HD~201601 and HD~24712 were observed in one and two ABBA sequence(s) with each exposure time of 60 sec, respectively.
For the telluric standards, A0V stars (HD~208108 and HD~15130) were also observed.
The GNIRS spectra are reduced using the Gemini IRAF package, and followed standard procedure, which includes flat-fielding, sky subtraction, wavelength calibration based on Ar-Xe lamp spectra, and extraction of 1D spectra.
The 1D spectra are normalized by continuum fitting.
Telluric absorption in the normalized target spectra is corrected by dividing by the normalized spectra of the A0V stars.

\begin{figure*}[th]
  \begin{center}
    \includegraphics[width=\linewidth]{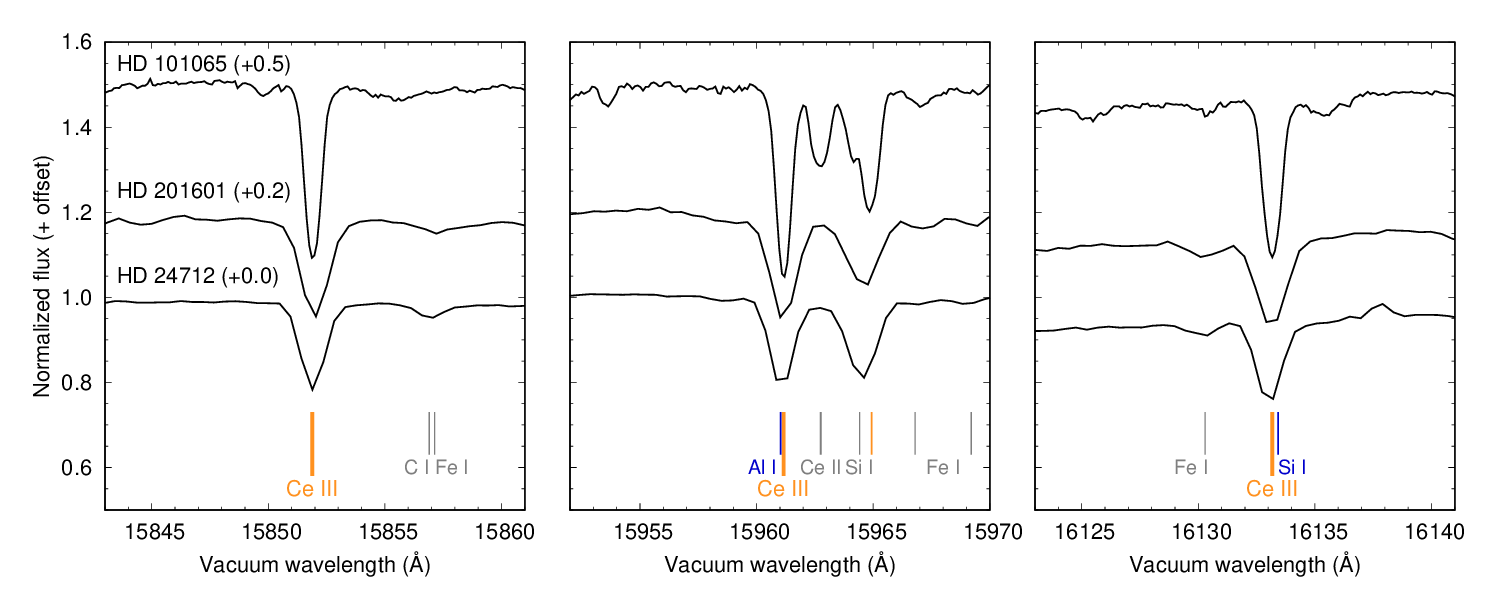}
\caption{
  \label{fig:Apspec}
  The observed spectra around the Ce III lines of the Ap stars.
  Absorption lines are labeled as in Figure \ref{fig:obs_spec}.
}
\end{center}
\end{figure*}

The spectra of three Ap stars around the region of the Ce III lines are show in Figure \ref{fig:Apspec}.
For those spectra, we measure EWs of the three Ce III lines by Gaussian fitting.
Then, we perform spectral synthesis as in Section \ref{sec:gf} but by changing the Ce abundance and by using the derived $gf$-values of the Ce III lines.
We use the stellar parameters of the Ap stars derived in literatures using optical spectra as summarized in Table \ref{tab:paramAp}.
Although there are the atmospheric parameters derived using more sophisticated models that consider stratification of elements for all the samples \citep{Shulyak2009, Shulyak2010, Shulyak2013}, we adopt those derived using 1D atmospheric models (ATLAS) that we use for spectral synthesis.
We measure EWs of the Ce III lines in model spectra by Gaussian fitting and compare with the EWs in the observed spectra.
Comparison is repeated by varying Ce abundances until the model EWs match with the EWs in the observed spectra.

We show the results of the Ce III abundances estimated using the three Ce III lines in Table \ref{tab:abunCe}.
It is seen that the line-by-line variation of the abundances is small in each star.
This fact demonstrates that the derived $gf$-values of the Ce III lines are reasonable.

Table \ref{tab:abunCe} also shows the Ce II and III abundances reported in literatures, which confirms clear REE anomaly.
It is worth noting that, although \citet{Hubrig2012} showed that the derived Ce III abundances in HD~201601 deviate about 2 dex depending on the lines, the abundances derived by using the three Ce III lines here are consistent with each other.
Also, this is the first measurement of the Ce III abundance for HD~24712.

\begin{deluxetable*}{lccccccccc}[ht]
\tablewidth{0pt}
\tablecaption{Adopted stellar parameters for the Ap stars.}
\label{tab:paramAp}
\tablehead{
Star   &  Model &  $\teff$  &      log $g$       &  [Fe/H]  &  $\xi$   &  $v\sin i$ &  [Al/H]  &  [Si/H]  &  Reference \\
          &            &   (K)      & (cm~s$^{-2}$)  &             &  (\kms) &  (\kms)     &             &             &
}
\startdata
HD~101065 & ATLAS   &  6600  &  4.20  &  $-$0.72  &  1.00  &  3.5         &  $-$1.11 &      0.11  &  \citet{Cowley2000}                \\
HD~201601 & ATLAS   &  7700  &  4.20  &       0.19  &  2.00  &  7.0$^a$  &      0.66  &      0.11   &  \citet{Ryabchikova1997a}     \\
HD~24712   & ATLAS   &  7250  &  4.20  &  $-$0.21  &  1.00  &  6.6$^a$  &      0.06  &  $-$0.18 &  \citet{Ryabchikova1997b}     \\
\enddata
\tablecomments{
All $v\sin i$ are smaller than spectral resolution ($R\sim45000$ or 17800) and does not affect results. \\
$^a$ Adopted from \citet{Sikora2019}.
}
\end{deluxetable*}
\begin{deluxetable*}{llccc}[ht]
\tablewidth{0pt}
\tablecaption{Ce abundances derived using Ce III or II lines in the Ap stars.}
\label{tab:abunCe}
\tablehead{
[Ce/H]  &   &   HD~101065 &  HD~201601 &  HD~24712
}
\startdata
Ce III    &   15851.880 {\AA}                &  4.11                   &  4.09                   &  3.86    \\
             &   15961.157 {\AA}                &  4.06                   &  4.04                   &  3.72    \\
             &   16133.170 {\AA}                &  4.16                   &  4.26                   &  3.83    \\ 
             &   Mean $\pm$ standard error &  4.11$\pm$0.03  &  4.13$\pm$0.07  &  3.78$\pm$0.03  \\
             &   Literatures                 &  4.76$^a$  &  3.46--5.36$^c$  &                  \\ \hline\hline
Ce II     &               &  2.93$^a$   &  1.26$^d$           &  1.42$^e$  \\
             &               &  3.07$^b$   &  &  \\
\enddata
\tablecomments{
$^a$ \citet{Shulyak2010}. $^b$ \citet{Cowley2000}. $^c$ \citet{Hubrig2012}. $^d$ \citet{Ryabchikova1997a}. $^e$ \citet{Ryabchikova1997b}.
}
\end{deluxetable*}

\bibliography{references}
\bibliographystyle{aasjournal}

\end{document}